# ANTICORRELATIONS AND SUBDIFFUSION IN FINANCIAL SYSTEMS

K.Staliunas

E-mail: Kestutis.Staliunas@PTB.DE

**Abstract**

Statistical dynamics of financial systems is investigated, based on a model of a randomly coupled equation system driven by a stochastic Langevin force. It is found that in the stable regime the noise power spectrum of the system is 1/f-like: $\propto w^{-3/2}$ ($w$ is the frequency), that the autocorrelation function of the increments of the variables (returns of prices), is negative: $\propto -t^{-3/2}$ ($t$ is the delay) and that the stochastic drift of the variables (prices, exchange rates) is subdiffusive: $\propto t^{1/2}$ ($t$ is the time). These dependences corresponds to those calculated from historical \$/EURO exchange rates. The model can be generalized to arbitrary stable, noise driven systems of randomly coupled components.

It is universally recognized that the temporal autocorrelations of returns of prices $C(t) = <dx(t) \cdot dx(t-t)>$ in finance markets are nonzero for short times $t$, less than approximately half an hour [1,2]. $dx(t)$ is the return of the price $x(t)$, defined as $dx(t) = (x(t+T) - x(t))/x(t)$. The precise form of the autocorrelation function is however not known. Also no theoretical model is suggested explaining nonzero autocorrelations of financial time series.

The aim of this letter is to derive the autocorrelation function from a model of a randomly coupled equation system driven by a stochastic Langevin force, and to calculate the particular form of the autocorrelation function. Our derivations show, that the short time autocorrelations are negative and obey a power law $C(t) \propto -t^g$ with the exponent $g \approx 1.5$. This result is checked by statistic analysis of the high frequency historical \$/EURO exchange rates.

Negative autocorrelations can be related with a subdiffusive stochastic drift of the corresponding variable $x(t)$. Under "subdiffusion" is understood that the variance of stochastically drifting variable increases as $<(x(t)-x(0))^2> \propto t^b$ with increasing time $t$, with $b<1$. For comparison, the variance of a position of Brownian particle increases linearly with time $b=1$, following the well known Wiener diffusion law.

The negative autocorrelations can be also related with a 1/f - like form of the power spectra: $S(w) \propto w^{-a}$ with $a<2$.

Recent investigations have shown that stochastic drift of invariant variables (e.g. of the phase of the order parameter, of the position of a localized structure or of a vortex) in stable, spatially extended, and noise driven nonlinear systems is in general subdiffusive, and that the power spectra are 1/f - like, with the exponent $a$ explicitly dependent on the dimensionality of the space [3]. Significant is that the spatially extended systems have infinitely many degrees of freedom, summation over which (or integration in the continuum limit) produces the above listed peculiarities of the stochastic distributions (subdiffusion, 1/f-like spectra, and anticorelations), absent in systems with small numbers of degrees of freedom. Motivated by the results from [3] a similar analysis is performed here for stable financial markets where the number of traders is large. In the limit of infinitely large number of traders one can consider the system as a continuum, and the dynamics of such system may resemble that of continuous spatially extended systems.

In the analysis in [3]:

1) one starts with a nonlinear evolution equation:

$$\frac{d\mathbf{A}}{dt} = \mathbf{N}(\mathbf{A}) + \Gamma(\mathbf{r},t) \tag{1}$$

for the temporal evolution of the state vector (order parameter) of the system:. $\mathbf{N}(\mathbf{A})$ is a deterministic, nonlinear operator acting on the state vector of the system, and $\Gamma(t)$ is the vector of $d$ - correlated noise terms $\Gamma(t) = (\Gamma_1(t), \Gamma_2(t), ..., \Gamma_j(t), ..., \Gamma_n(t))^T$, of strength $T$: $<\Gamma_j(t_1) \cdot \Gamma_k(t_2)> = 2T \cdot d_{jk} d(t_1 - t_2)$.

2) one linearizes around the stationary solution $\mathbf{A}_0$ of the equation (1) without the noise term: $\mathbf{A}(t) = \mathbf{A}_0 + \mathbf{a}(t)$, and obtains the linear equation system:

$$\frac{d\mathbf{a}(t)}{dt} = \mathbf{L} \cdot \mathbf{a}(t) \tag{2}$$

3) one diagonalizes (2) rewriting it in a new basis $\mathbf{a}(t) \to \mathbf{b}(t)$ corresponding to the eigenvectors of the matrix $\mathbf{L}$:

$$\frac{d\mathbf{b}(t)}{dt} = \Lambda \cdot \mathbf{b}(t) + \Gamma'(t) \qquad (3)$$

where $\Lambda$ is a diagonal matrix consisting of the eigenvalues of the matrix $\mathbf{L}$, and $\Gamma'(t)$ is the noise vector in the new coordinates.

4) one calculates the power spectra of the normal modes $\mathbf{b}(t)$ as driven by noise. After Fourier transform $\mathbf{b}(t) \to \mathbf{b}(w) = \int \mathbf{b}(t) \exp(iwt) dt$ one obtains the expressions for spectra of each perturbation mode: $b_j(w) = \Gamma'_j(w)/(iw + l_j)$.

5) one makes a summation (or integration in the limit of a continuum) of the spectra of normal modes to obtain the total power spectrum of the system:

$$S(w) = \sum_j S_j(w) = \sum_j |b_j(w)|^2 = \int r(l) |b(l,w)|^2 dl \qquad (4)$$

In the case of a continuum the density of distribution of eigenvalues $r(l)$ over the complex plane $l$ is significant.

We use a similar procedure for the analysis of stable financial markets, or at least of stable periods of generally unstable markets. We assume that a large number of traders are nonlinearly coupled one with another, which can be described generally by operator $\mathbf{N}(\mathbf{A},t)$, which, in general can be nonstationary (explicitly dependent on time $t$). The operator $\mathbf{N}(\mathbf{A},t)$ includes all the trading strategies of traders. It is also assumed that the financial market is subjected to random $d$-correlated perturbations, corresponding to political events and similar, described by external noise of particular strength $T$. Then the system under investigation can be formally described by (1).

If one knows the nonlinear operator $\mathbf{N}(\mathbf{A},t)$, and when the system is stationary and stable, the power spectra can be calculated following the above guidelines (1-4). The difficulty is that in principle one cannot know the particular form of the nonlinear operator for the general case of financial systems. Therefore we will not guess the form of nonlinear operator $\mathbf{N}(\mathbf{A})$ in (1), but, using assumptions as detailed below, we will guess the form of linear matrix $\mathbf{L}$ in (2). Indeed, if the system (1) in absence of random force has a stationary solution, then the linearization of the corresponding nonlinear operator $\mathbf{N}(\mathbf{A})$ is in principle possible in the vicinity of the this solution: $\mathbf{A}(t) = \mathbf{A}_0 + \mathbf{a}(t)$, where

$\mathbf{a}(t) = (a_1(t), a_2(t), ..., a_j(t), ..., a_n(t))^T$ stands for the perturbation of the state vector of the system (*n*, the number of the traders is large). Then (2) is also possible. Since the trading strategies of different traders may be very different, then it is plausible to assume that the coupling matrix **L** is a random matrix. The stability assumption imposes a condition that the coupling matrix **L** must be negative definite.

Therefore the very general assumptions, that: 1) the system is stationary; 2) the system is stable; 3) the traders have many different strategies; 4) the system is subjected to external random perturbations, lead to the model (2). Differently from the "deterministic" case (e.g. complex Ginzburg-Landau equation with noise as investigated in [3]), where the linear coupling matrix **L** is systematically derived from the known nonlinear model, for financial markets the matrix **L** should be "guessed" or introduced phenomenologically, due to the lack of the knowledge on the underlying processes. The most natural guess is that the matrix **L** is a random negative definite matrix.

The random matrix **L** is then to be diagonalized, which generates Lyapunov exponents distributed over the complex $l$-plane. The power spectra of the corresponding normal modes are to be calculated as driven by external noise. The summation over the spectra of the modes (or the integration in the continuum limit, taking into account the density of distributions of eigenvalues) leads eventually to the power spectrum of a particular variable of the system.

The most natural choice of the random, negative definite matrix is $\mathbf{L} = -\mathbf{M} \cdot \mathbf{M}^T$, where **M** is square matrix of real random numbers, and $\mathbf{M}^T$ is its transposed matrix. In this case **L** is a Hermitian, symmetric, negatively definite matrix, thus all its eigenvalues are real and negative. This choice leads to analytic results, however, restricts the class of systems to the Hermitian ones. The main results of this letter concerns the Hermitian systems, however also the nonhermitian systems are discussed, and it is shown, that the statistical properties of the system are not modified significantly in nonhermitian cases.

We note that the generalized Lotka - Volterra models, describing the power law statistics of the distribution of the returns [4], lead to Hermitian systems and to Hermitian linear coupling matrices respectively. The general form of Lotka – Volterra models:

$$\frac{dA_i(t)}{dt} = r_i A_i(t) - A_i(t) f(\overline{A}(t), t) + g(\overline{A}(t), t) \tag{5}$$

accounts for the growth $r_i A_i(t)$, and for the global coupling of the individual components $A_i(t)$. A global coupling occurs due to competition of the components for the limited common resources $-A_i(t) f(\overline{A}(t), t)$, and due to subsiding-like phenomena $g(\overline{A}(t), t)$, and depends on the (arbitrary weighted) average $\overline{A}(t)$ of the individual components $A_i(t)$. The linearisation of (5) with random individual growth rates, and with the global coupling schemes used in [4], leads to Hermitian, negatively definite, random matrices.

**Power spectra**

The density of the eigenvalues of the Hermitian random matrix $\mathbf{L}$ as defined above is:

$$r(l) = \frac{1}{2\pi\sigma^2}\sqrt{\frac{4\sigma^2 - l}{l}} \tag{6}$$

and does not depend on the statistical distributions of the elements of matrix $\mathbf{M}$. $\sigma^2/n$ is the variance of the elements of matrix $\mathbf{M}$ of size $(n \times n)$.

The response of each perturbation mode to external noise, as follows from (2) is:

$$b_j(w) = \frac{\Gamma_j(w)}{iw + l_j} \tag{7}$$

The power spectrum of each perturbation mode (assuming delta correlated noise of strength T: $\langle \Gamma_j(w) \cdot \Gamma_k^*(w)\rangle = T\delta_{j,k}$ ) is:

$$S_j(w) = |b_j(w)|^2 = \frac{T}{w^2 + l_j^2} \tag{8}$$

The power spectrum of the system is the sum of power spectra of each normal mode (8), $S(w) = \sum_j S_j(w)$, which in the limit of infinitely large number of traders can be replaced by the integral of power spectrum (8) over eigenvalues, accounting for the density function (6):

$$S(w) = \int r(l) \cdot S_l(w) \cdot dl = \frac{T}{w^{3/2}\sigma}\left(1 + (4\sigma^2/w)^{-2}\right)^{1/4} \sin\left(\frac{\arctan(4\sigma^2/w)}{2}\right) \tag{9}$$

which in the limit of small frequencies $w \ll 4\sigma^2$ simplifies to:

$$S(w) = \frac{T}{w^{3/2}\sigma\sqrt{2}} \tag{10}$$

which means a $1/f^a$ power spectrum with the exponent $a = 3/2$.

A numerical check of the spectra (9,10) was performed. Matrices $\mathbf{M}$ of normally distributed real random numbers of size $(1000 \times 1000)$ and with $\sigma = 1$ were generated numerically. The density of the eigenvalues of the corresponding Hermitian matrix (Fig.1.a) follows the power law $r(l) \propto |l|^{-1/2}$ up to cut-off value $|l_{cut-off}| \propto 4s^2$, and drops rapidly to zero beyond this value. The corresponding spectra in log-log scale, as calculated by summation of response functions (8) over the normal modes is shown in Fig.1.b. The region characterized by a 3/2 slope, as extending to $w_{max}$ approximately of order of unity is clearly visible in accordance with analytic results. For large frequencies $w > 1$ (in general for $w > 4s^2$), the slope is 2, which corresponds to the times less than the reaction time of the market, determined by the cut-off value of the distribution of eigenvalues (6)

The numerical check was also performed for the matrices of nongaussian distributed real random numbers (uniform and exponential distribution were used), which did lead to any sensible differences with the plots in Fig.1, and which suggests that the spectrum following from our above model is also valid for the components coupled randomly in a nongaussian way.

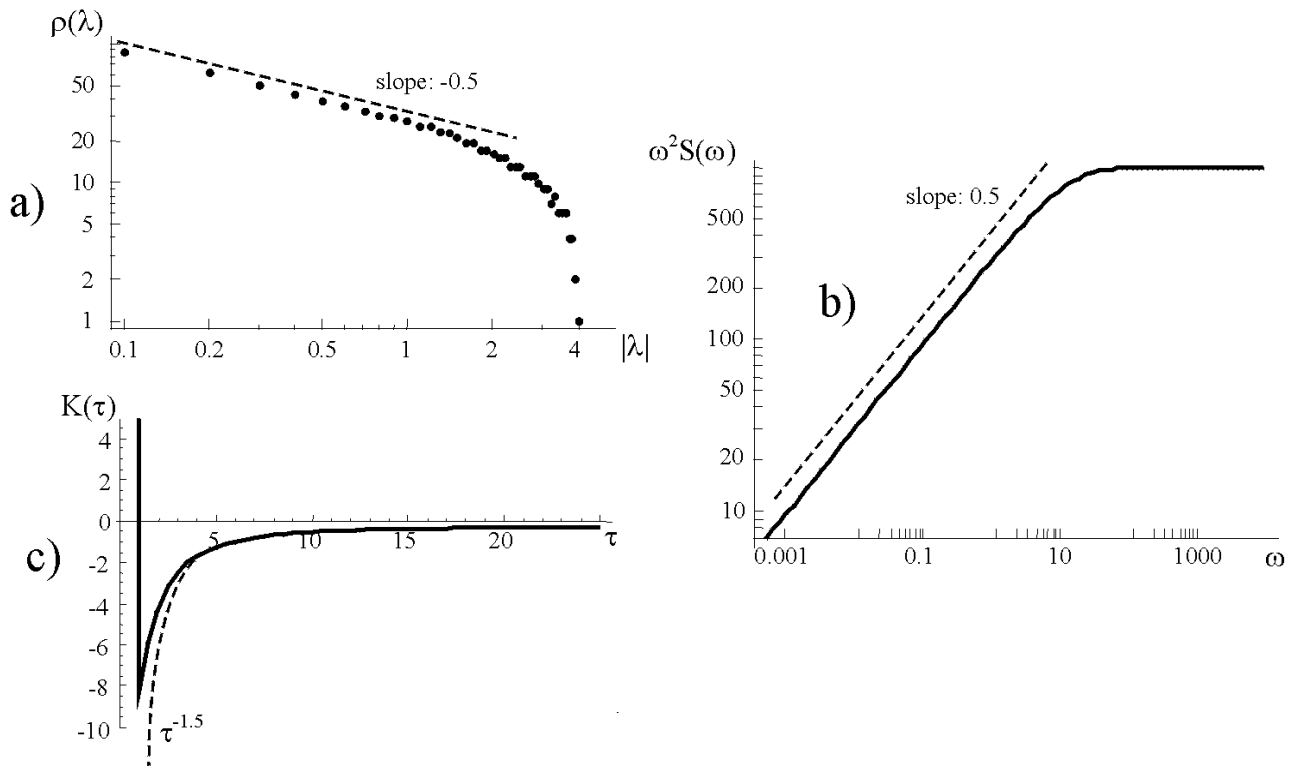

*Fig.1. a) Distribution of the eigenvalues of numerically generated Hermitian negatively definite random matrices of size (1000×1000) in log-log scale, as averaged over 10 realisations; b) the normalized power spectrum (multiplied by $w^2$), corresponding to power spectrum of the increments of variables; c) autocorrelation function of the increments of variables.*

**Stochastic drift**

The integral of the $1/f^a$ power spectrum diverges in the limit of small frequency for $a \geq 1$, which means that the variance of the increment of the variable grows to infinity for large times. The drift of the variable $\langle \Delta x(t)^2 \rangle = \langle (x(t) - x(0))^2 \rangle$ can be calculated from the power spectra: $\langle \Delta x(t)^2 \rangle \propto \int_{w_{min}}^{\infty} S(w) dw$, where $w_{min} \approx 2p/t$ is the lower cut-off boundary of the power spectrum (as follows from the Parseval theorem). Thus the drift of the variable obeys $\langle \Delta x(t)^2 \rangle \propto t^{a-1}$ for the power spectra $1/w^a$ as the direct integration yields. This generalizes the well known Wiener law for stochastic drift, stating that the variance grows linearly with time. From the results above it follows that the stochastic drift in financial time series should be weaker than Brownian. In particular the drift in the model described by a Hermitian random coupling matrix, ($a = 3/2$) is $\langle \Delta x(t)^2 \rangle \propto t^{1/2}$. The corresponding Hurst exponent [5] is then respectively $H = (a-1)/2$, and from our model of the Hermitian random coupling matrix $H = 1/4$.

We note that the calculated stochastic drift is valid for normal distributed external noise. If the external noise obeys stable Levy distribution (with the asymptotics $L(dx) \propto |dx|^{-1-m}$ for large $|dx|$, with $0 < m < 2$), then the Hurst exponent is respectively $H = a/2 + 1/m - 1$, as obtained by applying Parseval theorem for correlated Levy process (derivation details will be given elsewhere). In this case the competition between subdiffusion (due to anticorrelated signs of subsequent steps), and superdiffusion (due to leptokurtic Levy distributions of individual step sizes) occur. In particular for $m = 4/3$ the subdiffusion and superdiffusion mutually compensate and result in normal diffusion with $H = 1/2$.

**Autocorrelation functions**

The autocorrelation function of the increments can be calculated from the power spectra applying the Wiener-Kinchine theorem: $K(t) = \int S'(w) \exp(-iwt) dw$. Here $S'(w)$ stands for the power spectrum of the increments related with the power spectrum of the original variable by $S'(w) = w^2 S(w)$. Using (4) for power spectrum one obtains:

$$K(t) = \iint \frac{Tw^2}{l^2 + w^2} \exp(-iwt) r(l) dw dl = Td(t) + \int_{-\infty}^{0} Tl r(l) \exp(|t|l) dl \qquad (11)$$

which accounting for the density of eigenvalues described by (6) results in:

$$K(t) = Td(t) - T \cdot I_1(2s^2|t|)\exp(-2s^2|t|)/|t| \tag{12}$$

where $I_1$ is modified (hyperbolic) Bessel function of the first kind. The asymptotic of the autocorrelation function are: $K(t) = Td(t) - T\exp(-2s^2|t|)/2s^2$ for $|t| \leq 2s^2$, when the cut-off of the distribution (6) play a role, and $K(t) = Td(t) - T|t|^{-3/2}(2s^2)^{-1/2}$ for $|t| \geq 2s^2$, when the density of the eigenvalues follows asymptotically the power law $r(l) \propto l^{-1/2}$.

The autocorrelation function as obtained numerically is shown in Fig.1.c. It follows precisely the power law down to delays $|t| \approx 4$ in accordance with the analytical prediction (12).

**Nonhermitian case**

It may be anticipated that in the general case of stable financial systems (and in the general case of stable complex systems) the coupling matrix being negatively definite, is not necessary Hermitian. The nonhermiticity can appear in the frames of modified Lotka-Volterra models (5) if in addition to global coupling, the nonglobal coupling is introduced in (5), and/or if the global coupling in (5) acts with a temporal delay. In nonhermitian case the eigenvalues of matrix $\mathbf{L}$ are complex conjugated and distributed on the left (negative) half-plane of $l$. The power spectrum is then calculated by:

$$S_j(w) = |b_j(w)|^2 = \frac{T}{(w - l_{j,\text{Im}})^2 + l_{j,\text{Re}}^2} \tag{13}$$

We could not find a universal procedure of generating phenomenologically a nonhermitian negatively definite matrix, thus we could not determine uniquely the corresponding universal distribution of the eigenvalues on the complex plane $l$. We tried different choices for the nonhermitian negatively definite matrix: by generating it as $\mathbf{L} = -\mathbf{M}_1 \cdot \mathbf{M}_1^T + c(\mathbf{M}_2 - \mathbf{M}_2^T)$, where $\mathbf{M}_1, \mathbf{M}_2$ are square matrices of normally-distributed real numbers, and $c$ is nonhermiticity parameter. We also applied the natural selection criteria, i.e. selected the matrices with all negative eigenvalues from the randomly generated ensemble of square matrices of normally-distributed real numbers. In both cases the distributions of the eigenvalues are compatible with the asymptotical form $r(|l|) \propto |l|^{-c}$ on the left (negative) $l$-half-plane. The exponent $c$ varies from ½ (in Hermitian case) till 1 depending on the model.

The distribution of the eigenvalues $r(|l|) = |l|^{-c}$ yields the power spectrum of $S(w) \propto w^{-1-c}$ for $c < 1$, as direct integration shows. The stochastic drift obeys $\langle \Delta x(t)^2 \rangle \propto t^c$, and the autocorrelation function of the increments obeys: $K(t) \propto d(|t|) - |t|^{-2+c} \Gamma(2-c)$. In this way the main statistical properties of Hermitian case (1/f-like spectra, anticorrelations, and subdiffusion) hold in nonhermitian case too.

**Comparison with historical $/EURO exchange data**

The autocorrelation function and the dependence of stochastic drift, as following from the above model, were compared with those calculated for the historical high frequency $/EURO exchange rates [6]. The open circles in Fig.2.a show the autocorrelations of 2 min. returns. Although the anticorrelations for the delays of up to 30 minutes are visible, the correspondence with the power law autocorrelation as following from our theoretical model is not precise. One can nonetheless conclude on domination of anticorrelations (antipersistence) of price movements.

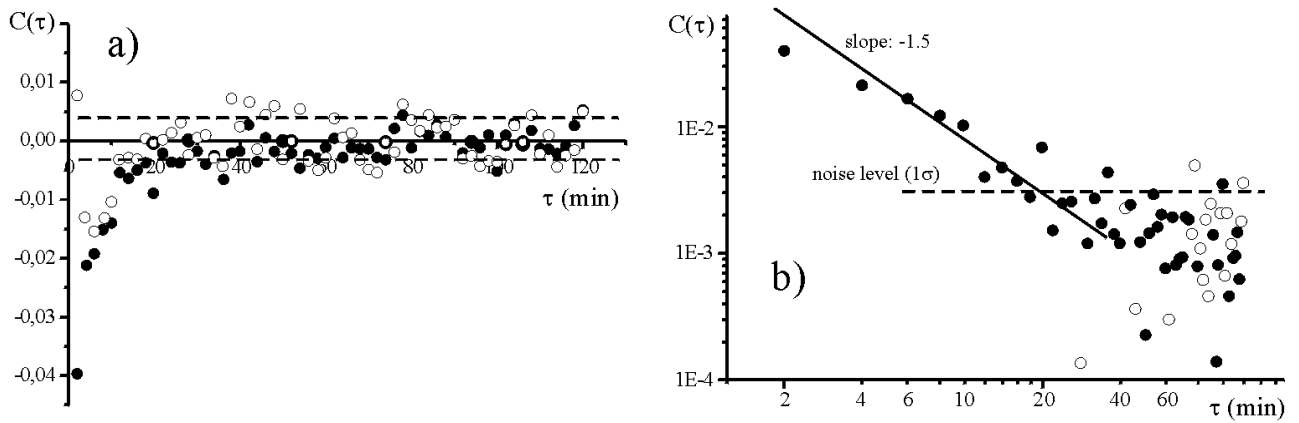

*Fig.2.* a) Autocorrelation $K(t) = \int dx(t) \cdot dx(t-t) \cdot dt \Big/ \int dx(t)^2 \cdot dt$ of the T=2 minute returns $dx(t) = (x(t+T) - x(t))/x(t)$ of the historical $/EURO exchange rates of the 2001 year (open circles), and of the signs of increments: $dx(t) = sign(x(t+T) - x(t))$ (solid circles) as a function of delay *t* in minutes. The dashed lines indicate the confidence range in 1*s* level.

b) Absolute values of autocorrelations of the signs of returns a function of delay *t* in log-log representation. Solid circles depict negative, and open circles – positive autocorrelations.

This lack of quantitative correspondence, to our opinion, is due to the fact that the dynamics of financial markets is in general nongaussian process. Therefore, in order to avoid a possible influence of the nongaussian statistics of the price variations we calculated autocorrelations of the signs of price increments $s(t) = dx(t)/|dx(t)|$. The results are given by the solid circles in Fig.2.a, which indicate that the anticorrelations of the signs of the price increments are more pronounced than the anticorrelations of the returns.

In Fig.2.b. the autocorrelations of the signs of the price increments are shown in log-log scale, where the solid (open) circles show the negative (positive) correlations. Although the data points are strongly scattered, the $K(t) \propto -|t|^{-g}$ dependence of the autocorrelation is plausible, with $g \approx 1.5$ for the time delays between 4 and 30 min.

The stochastic drift of $/EURO exchange rates was also calculated. The drift of the exchange rates was found only weakly subdiffusive: the local Hurst exponent was found slightly less than ½ in the time interval between 6 and 40 min. (open circles in Fig.3). and reached the minimum value $H_{min} \approx 0.44$. However the analysis of volatility normalized data, where the increments of the prices were substituted by the signs of the increments, show strong subdiffusive behaviour. The Hurst exponent is less than ½ up to times until 1 day, reaching the minimum value $H_{min} \approx 0.3$ in the time interval between 6 min. and 2 hours.

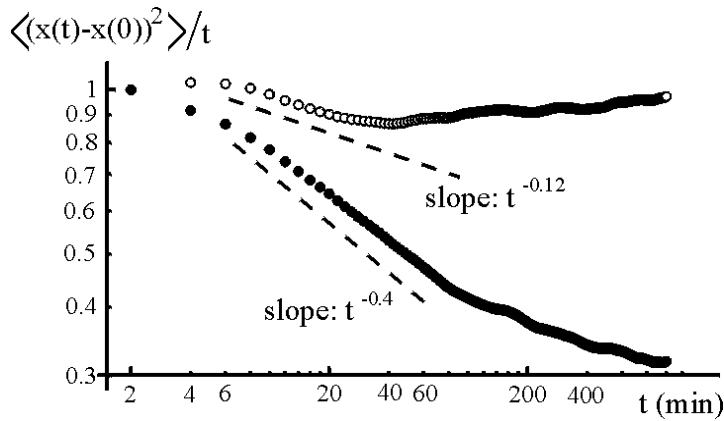

*Fig.3.* Open circles: stochastic drift of $/EURO exchange rate in log-log representation (the data as in Fig.2). Closed circles: stochastic drift of normalized to volatility exchange rates (the increments in the data are substituted by the signs of the increments). The dashed lines indicate the slopes –0.12 and –0.4 for the normalized stochastic drift, which correspond to drift exponents $b = 0.88$ and $b = 0.6$, and to Hurst exponents $H = 0.44$ and $H = 0.3$

**Conclusions**

The stochastic properties of the financial time series were calculated using a general model, assuming that: 1) large number of traders have different strategies; 2) external noise is present; 3) system is stationary, and 4) the system is stable. These qualitative assumptions lead to qualitative results, namely: to 1/f-like power spectra ($\propto w^{-3/2}$), to anticorrelated (antipersistent) dynamics of the increments ($\propto -t^{-3/2}$), and to subdiffusive stochastic drift of price ($\propto t^{1/2}$). The latter result is valid for normal distributed external noise, and is to be modified for the leptokurtic external noise.

The results following from the model correspond qualitatively with the results obtained from historical data [1,6], where the negative autocorrelations in particular interval of time delays are observed. The correspondence is better with the results obtained from the volatility normalized historical data: the exponents of autocorrelation function and of diffusion dependence match with those following from the model.

The main result (anticorrelations) following from our model can be compared with the real financial markets for sufficiently small time delays. E.g. the autocorrelations for the time delay of 1 hour, as following from our calculations ($K(t) \propto -|t|^{-3/2}$ dependence extrapolated to 1 hour, Fig.2.b.) should be of the order of $7*10^{-4}$. The length of historical data to check the autocorrelation of such magnitude with the certainty of $1\sigma$ level is approximately 12 years. Taking into account that the autocorrelations steadily decrease due to the globalisation of finance markets [1] the use of data from extremely long time intervals has no sense, thus the check of autocorrelations following from our model is difficult for delays larger than 30 min.

In the other limit of very small delays some inertia in finance markets may be present due to nonzero duration of trading operations (nowadays on the scale of 1 minute). Our model assumes the cutt-off of the density of the Lyapunov exponents, which rules out very fast dynamics, and which correspond to above discussed inertia of finance markets. In this limit (for the times delays $|t| \leq 2\sigma^2$ as follows from our model, and for delays less than approximately 4 min, as follows from historical data) the autocorrelation does not follow the power law, however remain negative. The autocorrelations of the signs of the increments calculated from historical data show the discrepance from the power law in accordance to our model. The autocorrelations of the returns however display small positive autocorelations for very small times ($\leq 2$ minute), which is however not explained by our model.

Our model is linear, and the stochastic Langevin forces are additive, which results in the Gaussian statistics of the returns, for the Gaussian distributed external noise. It is known that the multiplicative stochastic forces lead to nongaussian (Levy-, and truncated Levy-) distributions [3,7-9]. The corresponding modification of our model considering the multiplicative noise, is in progress.

The results from the paper apply not only to the financial markets, but in general to every noise driven system of randomly coupled components.

**References**

1. J-P. Bouchaud and M.Potters, Theory of Financial Risks, Cambridge University press, 2000.
2. P.Gopikrishnan, V.Plerou, L.A.N.Amaral, M.Meyer, and H.E.Stanley, Phys. Rev. E **60**, 5305 (1999);
3. K.Staliunas, Int. Journal of Bifurcation and Chaos **11**, (2001) 2845; K.Staliunas, Phys. Rev. E **64**, 066129 (2001);
4. S.Solomon, and M.Levy, J.Mod.Phys.C **7**, 745 (1996); O.Biham, O.Malcai, M.Levy, S.Solomon, Phys. Rev. E **58**, 1352 (1998);
5. H.E.Hurst, Trans.Am.Soc.Cin.Eng., **116** (1951) 292
6. Historical high frequency exchange $/EURO exchange rates of year 2001 were provided by INTERPRO Ltd.
7. D.Sornette and R.Cont, J.Phys. I France **7** (1997) 431
8. M.Marsili, S.Maslov and Y-C-Zhang, Physica A **253** (1998) 403
9. J.P.Bouchoud and M.Mezard, Physica A **282** (2000) 536